\documentstyle[prb,preprint,aps]{revtex}
\begin{document}
\draft
\tighten
\title{Theory of photoinduced charge transfer in weakly coupled
donor-acceptor conjugated polymers: application to an MEH-PPV:CN-PPV
pair}
\author{M. W. Wu and E. M. Conwell}
\address{Center for Photoinduced Charge Transfer, Chemistry Department,
University of Rochester, Rochester, New York 14627, USA\\
and Xerox Corporation, Wilson Center for Technology, 114-22D, Webster,
New York 14580, USA$^*$}
\maketitle

\begin{abstract}
In a pair of coupled donor-acceptor conjugated polymer chains,
it is possible
for an exciton photoexcited on either polymer
to decay into a hole in the
donor polymer's valence band and an electron
in the conduction band of the
acceptor polymer. We calculate the corresponding exciton decay rate
and its dependence on inter-polymer distance. For a pair of derivatives
of poly(phenylene vinylene), PPV, specifically
poly[2-methoxy, 5-(2$^\prime$-ethyl-hexyloxy)-1, 4 PPV], MEH-PPV,
and poly(2,5-hexyloxy $p$-phenylene cyanovinylene), CN-PPV,
at a separation of 6\ \AA ~the characteristic
decay time is $2.2$\ ps, whereas at 4\ \AA ~it is $\sim 50$\ fs.
\end{abstract}

\section{Introduction}

It was found
recently that photoconductivity can sometimes be greatly
enhanced by doping derivatives of
poly(phenylene vinylene), PPV,\cite{sar}
poly(thiophene)\cite{mo} and the $\sigma$-conjugated polymer
poly(methylphenylsilylene)\cite{ke} with the fullerene molecule
C$_{60}$. This led to construction of photovoltaic cells
containing a polymer-fullerene heterojunction.\cite{sar2,mo3,mo2,he}
However these devices had poor efficiency because few of the
photogenerated carriers were able to reach the
electrodes. Recently a great increase in efficiency was obtained by
the use of interpenetrating donor (D) and Acceptor (A) networks 
so that D-A junctions are distributed 
throughout the device.\cite{ha,yu}  In one embodiment 
D is MEH-PPV and A is CN-PPV.\cite{ha} 
Because of the cyano side groups on the PPV chains, the electron 
affinity of CN-PPV is greater than
that of MEH-PPV while the ionization potential of MEH-PPV is
less than that of CN-PPV. Therefore electrons in the excitons
photoexcited on MEH-PPV chains will tend to move onto CN-PPV
chains while holes in the excitons on CN-PPV chains tend to
move onto MEH-PPV.

Recently, Rice and Gartstein\cite{rice} 
developed a microscopic theory of the photogeneration process 
to calculate the rate of photoinduced
charge transfer due to exciton dissociation between a polymer
chain and an adjacent acceptor molecule. An essential feature
of this theory is that the transfer occurs with energy being
conserved, {\em i.e.}, no phonons are involved. They simplified
their theory by neglecting the effects of lattice relaxation,
{\em i.e.}, polaron formation.
Rice and Gartstein suggested that
their theory could be generalized to the case of the acceptor 
being another chain rather than a molecule. 
In this paper we carry out such a generalization to
study the dissociation of an exciton photoexcited on either the
donor or acceptor chain into a free hole in the
donor polymer's valence band and an electron
in the conduction band of the acceptor
polymer. The theory is applied to an MEH-PPV:CN-PPV pair.
We have extended it to investigate the dependence of the dissociation
rate on the distance between D and A.

\section{Model}

We start our discussion from a model Hamiltonian for
two coupled conjugated polymer chains of infinite length, $H=H_0
+H_1+H_2$. $H_0$ describes the intrachain and charge-transfer (CT)
electron-hole (e-h) pair excitations of the two
polymer chains considered independent.
By the term intrachain e-h excitations we refer
to the excitations of one chain without regard to the presence of the
other polymer chain, while by CT e-h excitation we mean the excitation
of a charge on one polymer chain in the presence of
a charge of opposite sign on the adjacent chain. Such excitations 
have also been called interchain excitons, indirect excitons and 
polaron pairs. $H_0$ includes
no coupling between the two chains except for the long-range
Coulomb interaction between electron and hole.
$H_1$ and $H_2$ represent the coupling between the
two chains, taken to be weak perturbations that cause the
interchain charge transfer. The former describes
single-particle hopping and the latter a process in which a
charge fluctuation in one species scatters a charge of the
other species from one polymer chain to the other.\cite{rice}

It is convenient to assume that each polymer chain has e-h
symmetry and measure energies from the center of its gap.
Because the gap centers of the two polymers need not coincide,
we will take our zero of energy at the gap center of the
second polymer, that of the first polymer differing from this
by the energy $\pm\Delta$. $H_0$ may then be written
\begin{eqnarray}
H_0&=&\sum_{n\sigma}[(\alpha_0\mp\Delta)a_{n\sigma}^\dagger a_{n\sigma}+
(\alpha_0\pm\Delta)b_{n\sigma}^\dagger
b_{n\sigma}-t(a_{n+1\sigma}^\dagger a_{n\sigma}
+\mbox{H.C.})\nonumber\\
&&\hspace{0.5cm}\mbox{}-t(b_{n+1\sigma}^\dagger b_{n\sigma}
+\mbox{H.C.})]-\sum_{mn\sigma s}U(n-m)b_{n\sigma}^\dagger
b_{n\sigma}a_{ms}^\dagger a_{ms}\nonumber\\
&&\mbox{}+\sum_{n\sigma}[{\bar \alpha}_0 (c_{n\sigma}^\dagger 
c_{n\sigma}+d_{n\sigma}^\dagger d_{n\sigma})-{\bar t}
(c_{n+1\sigma}^\dagger c_{n\sigma}
+\mbox{H.C.})\nonumber\\
&&\hspace{0.5cm}\mbox{}-{\bar t}(d_{n+1\sigma}^\dagger d_{n\sigma}
+\mbox{H.C.})]-\sum_{mn\sigma s}{\bar U}(n-m)
d_{n\sigma}^\dagger d_{n\sigma}c_{ms}^\dagger c_{ms}\nonumber\\
\label{h0}
&&\mbox{}-\sum_{mn\sigma s}V_1(n-m)
d_{n\sigma}^\dagger d_{n\sigma}a_{ms}^\dagger a_{ms}
-\sum_{mn\sigma s}V_2(n-m)
b_{n\sigma}^\dagger b_{n\sigma}c_{ms}^\dagger c_{ms}\;,
\end{eqnarray}
where $b_{n\sigma}^\dagger$ ($d_{n\sigma}^\dagger$) and
$a_{n\sigma}^\dagger$ ($c_{n\sigma}^\dagger$)
create an electron in the conduction band and a
hole in the valence band, respectively, located at the $n$th
monomer with spin $\sigma$ in the first (second) chain.
$\alpha_0$ ($\bar\alpha_0$) denotes the
center of each band measured from midgap.
The upper (lower) sign before $\Delta$ refers to the case
where the midgap is above (below) the zero of energy.
From the first and third terms, one can easily get the single
electron and single hole energy spectra: $\varepsilon_{ke}=
\alpha_0\pm\Delta-2t\cos(ka)$ and
$\varepsilon_{kh}=\alpha_0\mp\Delta-2t\cos(ka)$ of the first chain and
$\bar\varepsilon_{ke}
=\bar\varepsilon_{kh}=\bar\alpha_0-2\bar t\cos(k\bar a)$ of
the second chain.  Accordingly in these expressions $W=4t$
($\bar W=4\bar t$) is the band width. $a$ ($\bar a$) represents
the length of a monomer of the
first (second) polymer chain.
From the excitation spectrum, one can easily get the
energy gaps of e-h excitation, {\em ie.}, the minimum energy 
required to  create an e-h pair in one polymer chain, $E_g=2
(\alpha_0-2t)$ of the first chain and  $\bar E_g=2
(\bar \alpha_0-2\bar t)$ of the second chain. It is then 
straightforward to see that one should assume $\alpha_0>2t$
and $\bar\alpha_0>2\bar t$ in order to have $E_g$ ($\bar E_g$) $>0$.
The second and
fourth terms in Eq.\ (\ref{h0}) represent the intrachain 
Coulomb interaction with 
\begin{equation}
\bar U(n-m)=\bar U\delta_{nm}+\frac{\bar V}{|n-m|}(1-\delta
_{nm})\;,
\end{equation}
for the second chain and the corresponding expression for the
first chain with $\bar U$ and $\bar V$ replaced by $U$ and $V$.
The last two terms of Eq.\ (\ref{h0}) denote the interchain
Coulomb interactions. For simplicity, we assume $V_1(n-m)$
equals $V_2(n-m)$, with
\begin{equation}
\label{v}
V_1(n-m)=V_2(n-m)\equiv V(n-m)=\frac{V_1}{\sqrt{(n-m)^2+B}}\;,
\end{equation}
where $B=\kappa_\Vert d^2/(\kappa_\perp a^2)$ and $d$
stands for the perpendicular distance between 
the two chains.\cite{land}
$\kappa_\Vert$ and $\kappa_\perp$ are dielectric constants along
and perpendicular to the polymer chains, respectively.

The Hamiltonian
\begin{equation}
\label{h1}
H_1=t_\perp^\prime\sum_{n\sigma}(d_{n\sigma}^\dagger b_{n\sigma}+
\mbox{H.C.})+t_\perp^{\prime\prime}
\sum_{n\sigma}(c_{n\sigma}^\dagger a_{n\sigma}+\mbox{H.C.})
\end{equation}
represents the perturbation leading to the single-particle interchain
hopping process with $t_\perp^\prime$
and $t_\perp^{\prime\prime}$ being
electronic hopping integrals. In general, $t_\perp^\prime$
and $t_\perp^{\prime\prime}$ are not necessarily the same.
It is reasonable to assume they fall off exponentially
according to\cite{miz}
\begin{equation}
t_\perp^\prime=t_0^\prime e^{-\mu d}=t_0^\prime \exp (-\mu a\sqrt{B
\kappa_\perp/\kappa_\Vert})\;,
\hspace{0.3cm}\mbox{and}\hspace{0.3cm}
t_\perp^{\prime\prime}=t_0^{\prime\prime} e^{-\mu d}
=t_0^{\prime\prime}\exp (-\mu a\sqrt{B\kappa_\perp/\kappa_\Vert})\;.
\end{equation}
The charge-fluctuation transfer process is described by
\begin{equation}
\label{h2}
H_2=\sum_{mn\sigma s}v_1(n-m) (b_{n\sigma}^\dagger
c_{ms}^\dagger c_{ms}d_{n\sigma}+\mbox{H.C.})
+\sum_{mn\sigma s}v_2(n-m) (a_{n\sigma}^\dagger
d_{ms}^\dagger d_{ms}c_{n\sigma}+\mbox{H.C.})\;,
\end{equation}
where the first term corresponds to an electron being scattered
to the other chain by a local fluctuation of the hole density
in the vicinity of site $m$, and in the second term the roles of electron
and hole are reversed.
We further assume that the matrix element $v_{1(2)}(n-m)
={\tilde v}_{1(2)}/\sqrt{(n-m)^2+B}$, because, like $V_1(n-m)$ and
$V_2(n-m)$, it depends on the Coulomb attraction between the electron
and hole on separate chains.
It is noted that $\tilde v_{1(2)}$ is also related to the
electron hopping integrals and would vanish if $t_\perp^\prime$
($t_\perp^{\prime\prime})=0$. Therefore, we further assume
$\tilde v_{1(2)}=v_{1(2)} e^{-\mu d}$. However,
we stress here that $v_1$ ($v_2$) may also have additional dependence 
on $B$. Again, $v_1$ is not necessarily equal to $v_2$.

In the following discussion of charge transfer,
we always assume that the exciton is
photogenerated on the second chain. The exciton can dissociate
either with transfer of an electron or a hole to the first chain
as shown schematically in Fig.\ 1. It can also be deduced
from Fig.\ 1 that only for the midgap of the first chain above
that of the second chain is it possible to have hole
transfer (HT) without electron transfer, while only for
the midgap of the first chain below that of the second chain is
it possible to have electron transfer (ET) without
hole transfer. Additionally, for both HT and ET, in order that
the other carrier not hop to the other chain as well, it is necessary
to satisfy the additional condition $\bar E_g<E_g+2\Delta$.

We first discuss in detail
the case of hole transfer. With the intrachain and
interchain Coulomb interaction included, the intrachain e-h
excitation spectrum of the second polymer chain
consists of an intrachain exciton with energy $\bar\omega_0$
and a continuum band of width $2\bar W$ above the energy gap
$\bar E_{g}$. The HT e-h excitation spectrum consists of an
interchain (HT) exciton with energy $\omega_{HT}$ and
a continuum band of width $\bar W+W$ above the gap $\bar E_g/2
+E_g/2-\Delta$. (See Fig.\ 1). Energy conservation requires that CT
due to an intrachain exciton in the second chain giving up
a hole to the first chain (HT) can only take place when
\begin{equation}
\label{ht}
\bar E_g/2+E_g/2-\Delta\leq\bar\omega_0\leq\bar E_g/2+
E_g/2-\Delta+\bar W+W\;.
\end{equation}

The decay rate of the exciton can be written from the Fermi
golden rule:
\begin{equation}
1/\tau=(2\pi/\hbar)\sum_{\alpha f}|M_{\alpha f}|^2\delta(\omega_0-
\varepsilon_f)\;.
\end{equation}
In this equation, $M_{\alpha f}=\langle f|
H_\alpha|i\rangle$ ($\alpha=1$, 2)
is the matrix element describing the transition from the initial 
intrachain exciton state $|i\rangle$ to the final CT e-h excitation
states.

The intrachain spectrum can be acquired as follows. We first
construct a two-particle wave function of a singlet e-h pair excitation
on the second polymer chain as
\begin{equation}
\bar\Phi=\frac{1}{\sqrt{2}}\sum_{nm\sigma}\bar\Phi_{nm}
c_{n\sigma}^\dagger d_{m -\sigma}^\dagger |0\rangle\;,
\end{equation}
with $|0\rangle$ denoting the ground state. $\bar \Phi_{nm}$ is assumed
to be real and is normalized according to $\sum_{nm}\bar \Phi_{nm}^2
=1$. From $H_0\bar \Phi=E\bar\Phi$, we obtain for the lowest excitonic
bound state with energy $\bar\omega_0$,\cite{rice}
\begin{equation}
\label{eq1}
(E-2\bar\alpha_0)\bar\Phi_{n-m}=-2\bar t(\bar\Phi_{n-m+1}+\bar
\Phi_{n-m-1})-\bar U(n-m)\bar\Phi_{n-m}\;.
\end{equation}
Similarly we can construct a two-particle HT wavefunction as
\begin{equation}
\label{phih}
\Phi_{HT}=\frac{1}{\sqrt{2}}\sum_{nm\sigma}\phi_{nm}
c_{n\sigma}^\dagger b_{m-\sigma}^\dagger |0\rangle\;.
\end{equation}
 $\phi_{nm}$ is given by the equation
\begin{equation}
\label{eqb}
[E-\bar\alpha_0-(\alpha_0-\Delta)]\phi_{n-m}=
-(t+\bar t)(\phi_{n-m+1}+\phi_{n-m-1})-V(n-m)\phi_{n-m}\;,
\end{equation}
with the normalization condition $\sum_{n}{\phi
_{n}}^2=1/N$, where $N$ is the number of monomers on the chain.
 (here we have assumed $\phi_{n}$ to be real).
It is noted that in the above we only retain the solutions with zero
CM wave vector, indicated by a superscript $0$ on $\Phi$
and $\bar\Phi$, as is required for momentum conservation.
Because we assume that free carriers are generated by the dissociation,
we only need to get the scattering states of these
equations in order to calculate $1/\tau$.

The decay rate of an exciton in the second chain by transferring
a hole to the first chain may now be written as
\begin{equation}
\label{decay}
1/\tau_\alpha=(2\pi/\hbar)(g_{\alpha}^2/\bar E_g)
\Gamma_\alpha
\end{equation}
with $\alpha=1$ standing for the process of single-particle
hopping and $\alpha=2$ for the charge fluctuation process. We take
$g_{1}=t_0^{\prime\prime}$ and $g_{2}=v_2$.
The dimensionless rate $\Gamma_\alpha$, which differs from that
defined by Rice and Gartstein, is defined by
\begin{equation}
\label{gamma}
\Gamma_\alpha=\bar E_g\sum_\nu |M_\alpha(\nu)|^2
\delta(\bar\omega_0-\varepsilon_\nu)\;,
\end{equation}
where
\begin{eqnarray}
\label{m1}
M_1(\nu)&=&N\sum_n e^{-\mu a\sqrt{B\kappa_\perp/
\kappa_\Vert}} \phi_n(\nu)\Phi_n^0\\
\label{m2}
M_2(\nu)&=&N\sum_n\frac{e^{-\mu a\sqrt{B\kappa_\perp
/\kappa_\Vert}}}{\sqrt{n^2+B}}\phi_n(\nu)\Phi_n^0\;.
\end{eqnarray}

\section{calculations and Discussions}

We apply our theory to an MEH-PPV:CN-PPV pair. We consider
first an exciton photoinduced on the CN-PPV chain. This exciton is
dissociated by transferring a hole to MEH-PPV. To this end, we
take the second chain as CN-PPV and the first one as MEH-PPV.
The values of the various unknown parameters in Eq.\ (\ref{eq1})
(written with unbarred quantities for the first chain) must be
chosen to give the correct gap and exciton binding energy. For
MEH-PPV the optical absorption edge is at 2.1\ eV.\cite{yu1}
The single particle energy gap of MEH-PPV has been measured as
2.45\ eV.\cite{cam} The exciton binding energy is the difference
of these two numbers,\cite{conwell} thus 0.35\ eV. The intrachain
Coulomb potential coefficient $V$ or $\bar V$ is of the order of
the Coulomb attraction between an electron and a hole separated
by one monomer ($a=6.5$\ \AA), which is $\sim 1/2$\ eV.\cite{rice}
$V\sim 0.42$\ eV. Choosing $\alpha_0=3.2$\ eV, $t=U=1$\ eV,\cite{rice}
and $V=0.44$\ eV, we obtain from Eq.\ (\ref{eq1})
$E_g=2.4$\ eV, the exciton creation energy
$\omega_0=2.065$\ eV and its binding energy $\epsilon_b=0.335$\ eV.
For CN-PPV the optical absorption edge is at $\sim
2.3$\ eV.\cite{sam} It is reasonable to assume that the
exciton binding energy is similar to that of PPV and MEH-PPV. With
$\bar\alpha_0=3.3$\ eV, $\bar t=\bar U=1$\ eV,
and $\bar V=0.4$\ eV, we obtain $\bar E_g=2.6$\ eV, the exciton
creation energy $\bar \omega_0=2.285$\ eV and
$\bar\epsilon_b=0.315$\ eV.

To obtain numerical results for HT we further take
$\kappa_\Vert=8$,\cite{fiok} $\kappa_\perp=3$,
$V_1=0.44$\ eV, the energy difference between gap centers of
MEH-PPV and CN-PPV $\Delta=0.5$\ eV,\cite{ha} and
$\mu=1.18$\ \AA$^{-1}$.\cite{miz}
For the numerical calculations, we employ a chain of $N=400$ unit
cells for each of the polymers and use periodic boundary conditions.
We then numerically solve Eq.\ (\ref{eqb}) for the CT band
eigenfunctions $\phi_n(\nu)$ and their eigenvalues $\varepsilon_\nu$
and substitute them and the numerical solution of Eq.\ (\ref{eq1})
into Eq.\ (\ref{gamma}) to get $\Gamma_\alpha$. In
evaluating $\Gamma_\alpha$, we replace
the $\delta$-function in Eq.\ (\ref{gamma}) by a Gaussian with
variance $\sigma=0.032$\ eV. The resulting $\Gamma_\alpha$
($\alpha=1$, 2) are plotted as solid curves as a function of
$B$. Since $B$ is proportional to the square of the distance between
the chains, the transfer rate falls off very rapidly with increasing
$B$. We estimate the minimum chain separation as $\sim 4$\ \AA,
giving the minimum $B\sim 1$. An increase in interchain distance
by a factor 2 is seen to decrease the transfer rate $\Gamma_1$ by
almost 4 orders of magnitude, and $\Gamma_2$ even more.

Rewriting the first inequality in Eq.\ (\ref{ht}) we obtain the
threshold value of $\Delta$,
\begin{equation}
\label{cc}
\Delta_T=E_g/2-\bar E_g/2+\bar\epsilon_b\;.
\end{equation}
For HT in this case  Eq.\ (\ref{cc}) gives $\Delta_T=0.2$\ eV.
It is noteworthy that $\Gamma_1$ and $\Gamma_2$ peak at the
threshold value of $\Delta$. This was also found to be the case
for the calculation of charge transfer to a dopant
molecule.\cite{rice}  The appearance in Fig.\ 2 of finite $\Gamma$
values below the threshold is due to the replacement of the
$\delta$-function in Eq.\ (\ref{gamma}) by a Gaussian.

It is possible to change $\Delta$ by changing one or both of the
polymers or by application of an electric field perpendicular
to the chains. For an electric field of $5\times 10^6$\ V/cm,
for example, if the MEH-PPV and CN-PPV chains were separated by 4\ \AA,
~the change in $\Delta$ due to the field would be $\pm 0.2$\ eV.
Thus $\Delta$ for HT in MEH-PPV:CN-PPV would be either 0.3 or 0.7\ eV
depending on the direction of the electric field. As can be seen in
Figs.\ 3(a) and 3(b), for $\Gamma_1$ this could be a change by a
factor 2.  For $\Gamma_2$ the change is smaller because $\Gamma_2$
decreases less rapidly with increase in $\Delta$. Thus the dissociation
rate may be substantially increased or decreased
by high electric fields.

Having $\Gamma_1$, we can get an approximate value, or at least
an upper limit, for $\tau_1$ using the average of $t_0^{\prime\prime}$
values for MEH-PPV and CN-PPV, the individual values having been
determined from the energy shift between exciton and excimer 
emission.\cite{wu} That average is $t_0^{\prime\prime}=24$\ eV.
For a spacing of 6\ \AA ~between MEH-PPV and CN-PPV $B=2.25$, leading to
$\Gamma_1=2\times 10^{-7}$ for $\Delta=0.5$\ eV and $\tau_1=2.17$\ ps.
For 4\ \AA, $B=1.00$ leading to $\tau_1=47.7$\ fs for $\Delta=0.5$\ eV.
Thus the dissociation process
we are calculating is quite rapid.

Consider now the case of an exciton photoinduced in MEH-PPV
dissociating into an e-h pair by transferring an electron to CN-PPV.
For this calculation we take the second chain to
be MEH-PPV, thus interchanging $\bar E_g$, $\bar\alpha_0$, $\bar t$,
$\bar U$ and $\bar V$ with $E_g$, $\alpha_0$, $t$, $U$
and $V$ respectively. Equations (\ref{ht}) and (\ref{eqb})
derived for the HT case are unchanged. Numerical solution of
the equations, as described earlier, leads to $\Gamma_\alpha$
for ET shown as the dashed curves in Fig.\ 2 and 3. The
threshold value of $\Delta$ for ET determined from Eq.\ (\ref{cc})
is 0.4\ eV rather than 0.2\ eV because of the interchange of $E_g$ and
$\bar E_g$; the exciton is generated on the chain with the smaller
gap in this case. Apart from the difference in
threshold, $\Gamma_\alpha$ for ET varies with $\Delta$ and $B$ in
similar fashion to the case for HT. The decay rate $1/\tau_\alpha$ is
obtained from Eq.\ (\ref{decay}) with $g_1=t_0^\prime$, $g_2=v_1$.
Since $t_0^\prime=t_0^{\prime\prime}$ and $\Gamma_1$ is about
twice as large for ET when $\Delta=0.5$\ eV,
the dissociation rate $1/\tau_1$ is about
twice as large for ET as for HT 
and the $\tau_1$ values 1/2 as large.

In summary we have calculated the dissociation rate of an exciton
on a donor or acceptor chain into a free electron on A and a free
hole on D by a process that conserves energy, {\em i.e.}, no phonons
are involved. Also, the process is applicable to chains that are
sufficiently long that broadening of their molecular orbital levels
is at least as large as the spacing of the unperturbed discrete
levels.\cite{rice} For dissociation to be possible in the
absence of an electric field whichever chain
the exciton is generated on, the midgap of D must lie above
that of A by at least the exciton binding energy plus $1/2$ the
absolute value of the difference in the energy gaps. For the
difference in midgaps greater than this amount the
rate of dissociation decreases rapidly. The difference in midgap
energy levels of the pair CN-PPV:MEH-PPV is close to satisfying
the condition for maximum transfer rate. The transfer rate falls off
so rapidly as  the distance between D and A increases that
essentially transfer can take place only between neighbors
separated by perhaps 4 to 6\ \AA ~perpendicular to the chain
direction. For a CN-PPV:MEH-PPV pair separated by 4\ \AA ~we
calculate a hole transfer time $\sim 50$\ fs, the electron
transfer time being shorter by a factor 2. As a consequence, since
it has been demonstrated that there is phase separation of the
two polymers on a scale of 10-100\ nm,\cite{ha} and the exciton
hopping time in PPV has been estimated as 0.4\ ps,\cite{baa} the
dissociation rate will likely be determined by the time it takes
the exciton to diffuse to a location where the required donor or
acceptor is within $\sim 4$\ \AA.

We acknowledge the support
of the National Science Foundation
under Science and Technology Center grant CHE912001.

\begin{figure}
\caption{Schematic of the relations between band structures
required for hole transfer (center and right columns) and electron
transfer (center and left columns) with exciton created on chain 2.
The dashed line across the three columns is the zero of the
energy scale chosen to coincide with midgap on chain 2. Midgaps of
the other two chains are indicated with dashed lines.}
\end{figure}

\begin{figure}
\caption{$\Gamma_\alpha$ ($\alpha=1$, 2) as a function of $B$,
where $B=0.063d^2$, $d$ being the perpendicular distance between chains
in \AA.
Hole transfer: solid curves; Electron transfer: dashed curves.}
\end{figure}

\begin{figure}
\caption{(a): $\Gamma_1$ as a function of $\Delta$
for different values of $B$.  (b):  $\Gamma_2$ {\em vs.} $\Delta$
 for different values of $B$.}
\end{figure}

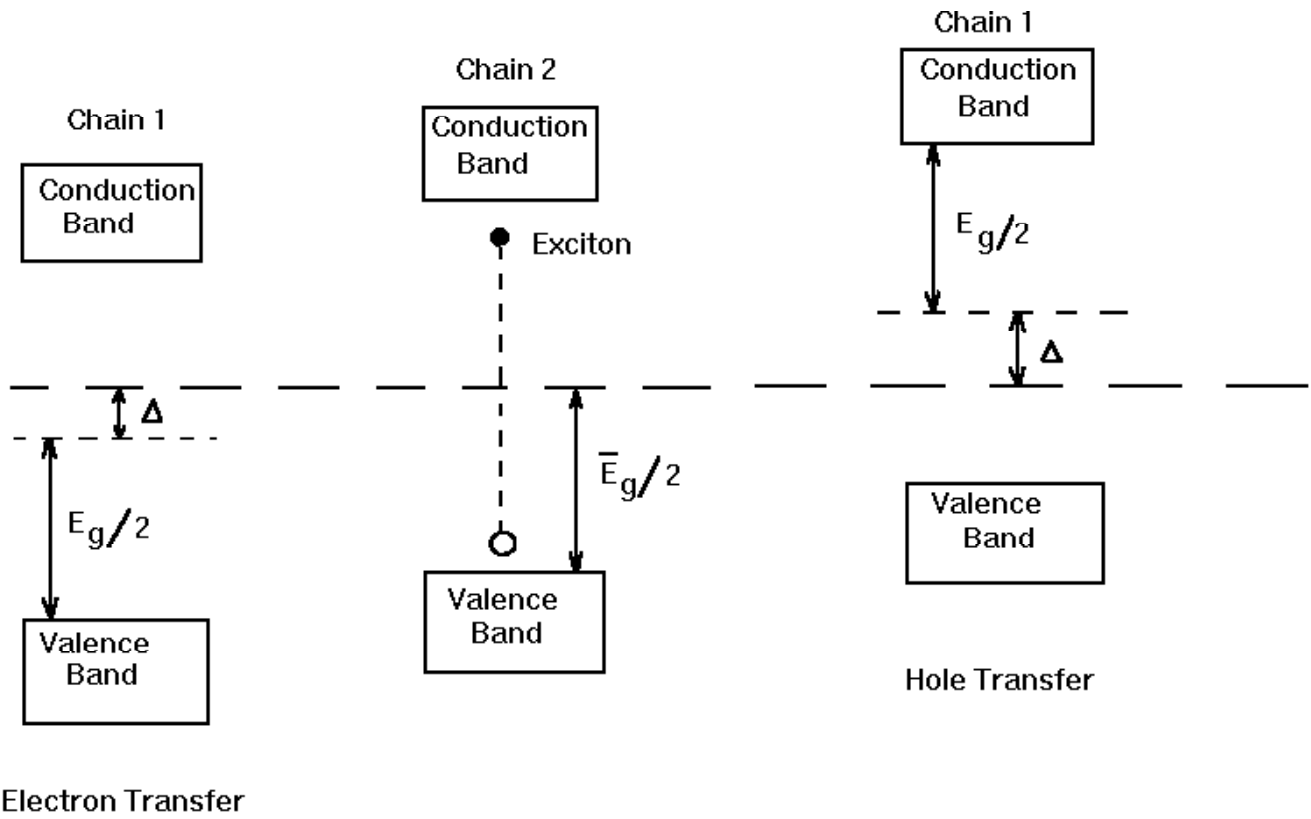
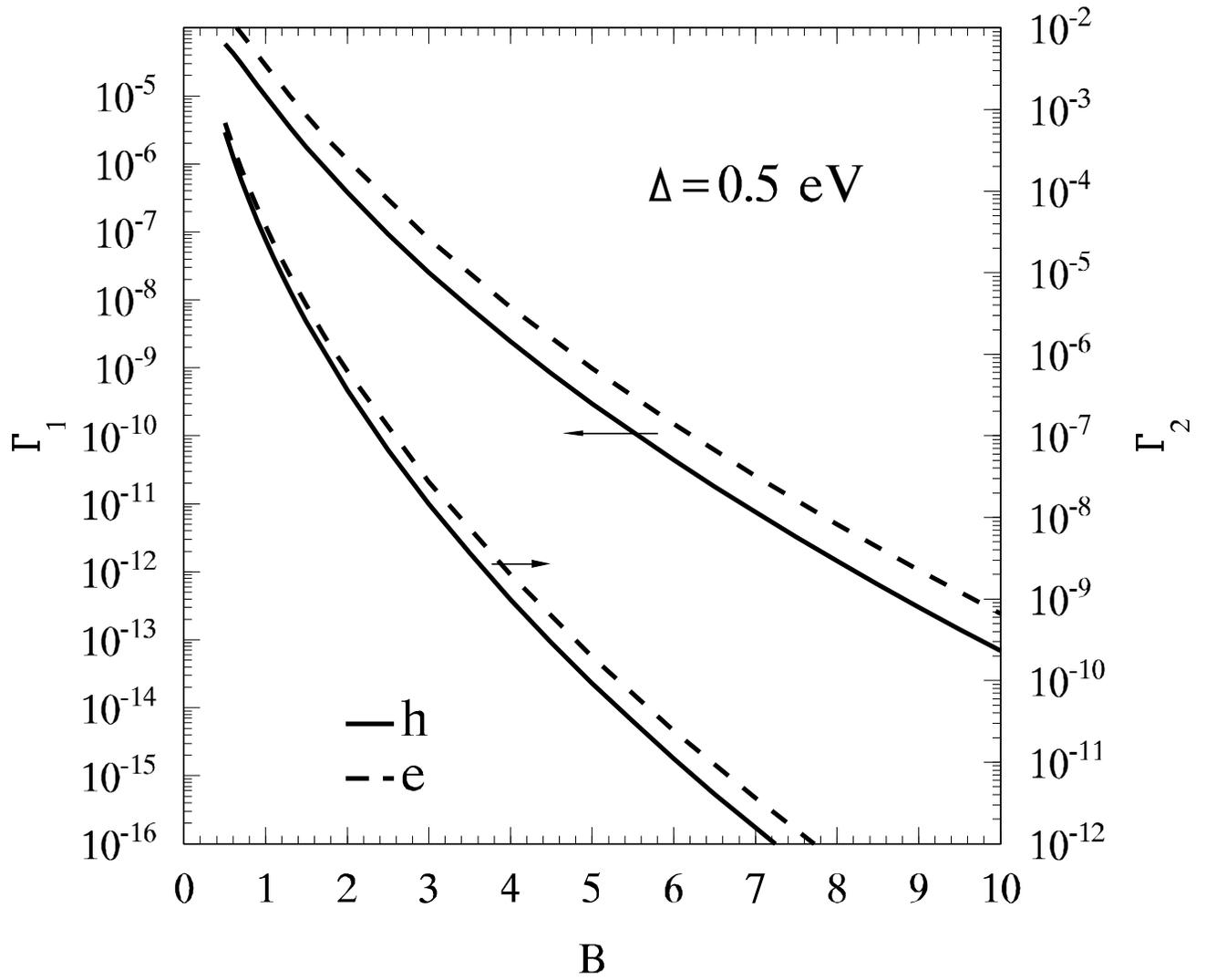
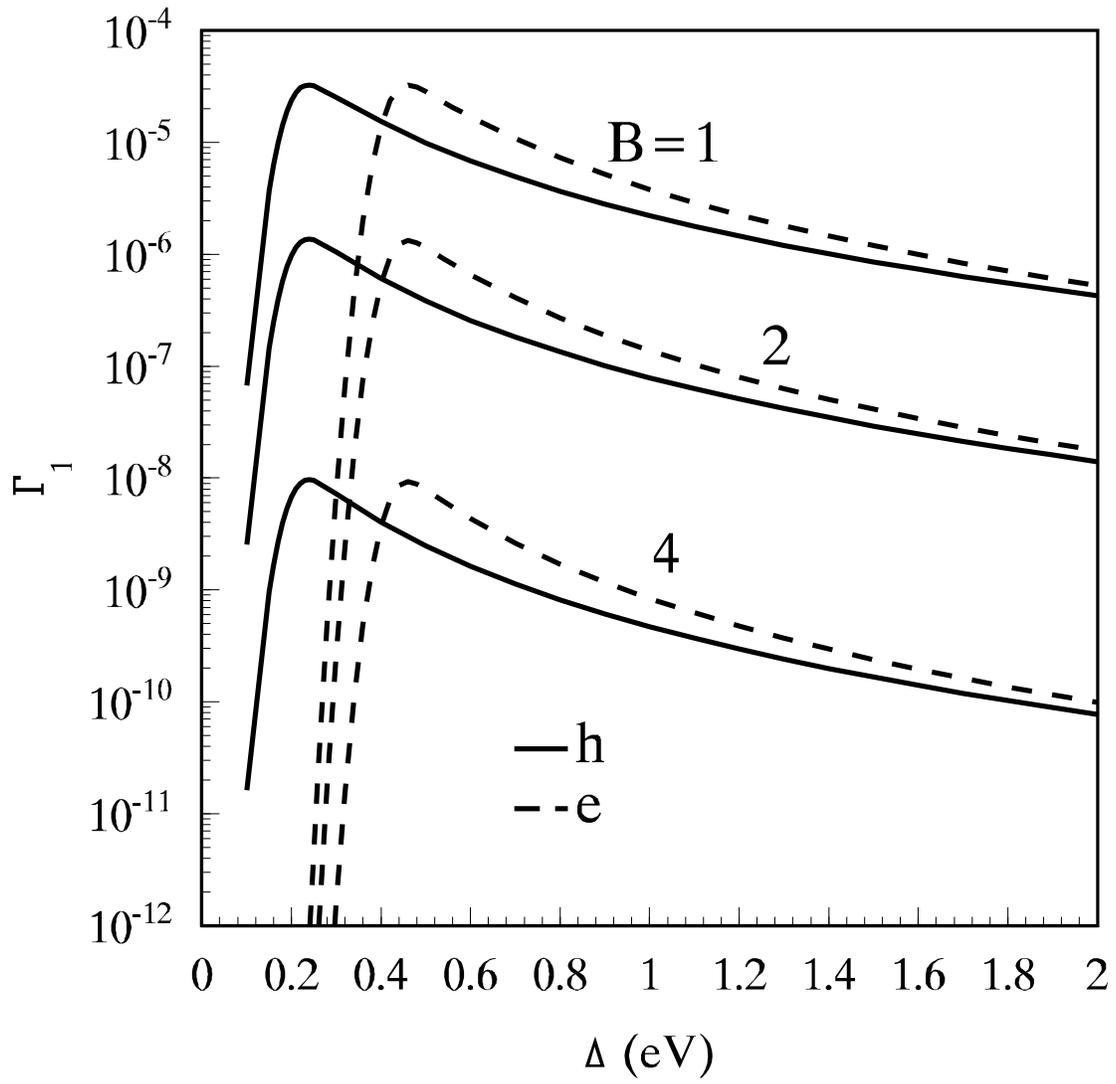
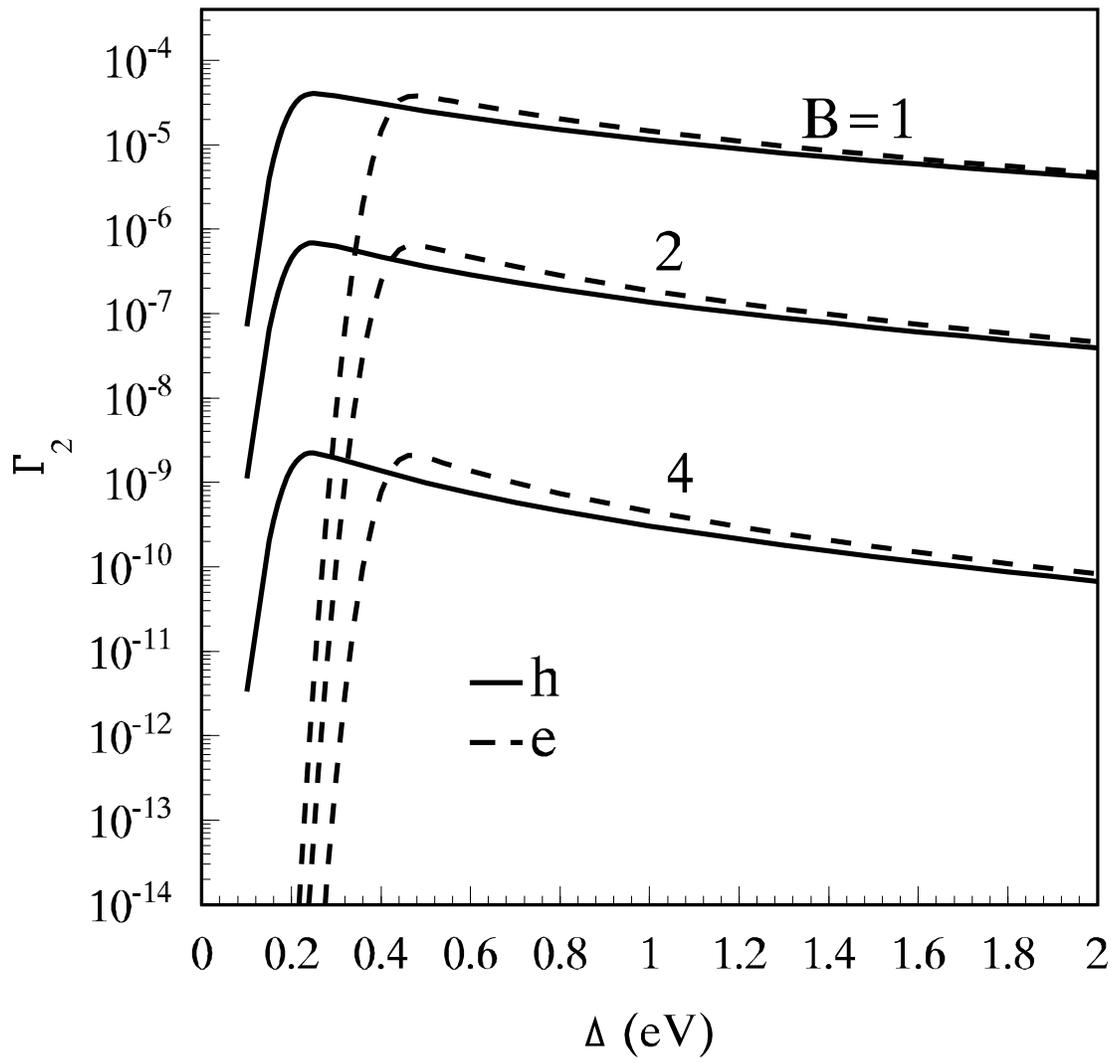
\end{document}